\documentclass[journal]{IEEEtran}

\usepackage{times}
\usepackage{epsfig}
\usepackage{graphicx}
\usepackage{amsmath}
\usepackage{amssymb}
\usepackage{graphics}
\usepackage{multirow}
\usepackage{siunitx}
\usepackage{url}
\ifCLASSINFOpdf
\else
\fi
\usepackage{xcolor}

\begin{document}
%
\title{Joint Demosaicing and Super-Resolution (JDSR): Network Design and Perceptual Optimization}
%
%
%

\author{Xuan~Xu,
        Yanfang Ye, and Xin~Li,~\IEEEmembership{Fellow,~IEEE}
\thanks{X. Xu and X. Li are with the Lane Department
of Computer Science and Electrical Engineering, West Virginia University, Morgantown,
WV, 26505 USA. Yanfang Ye is with Department of Computer and Data Sciences, Case Western Reserve University,
Cleveland, OH 44106. E-mail: xuxu@mix.wvu.edu, xin.li@mail.wvu.edu, yanfang.ye@case.edu}
}

%
%

\markboth{Journal of \LaTeX\ Class Files,~Vol.~14, No.~8, August~2015}%
{Shell \MakeLowercase{\textit{et al.}}: Bare Demo of IEEEtran.cls for IEEE Journals}
%



\maketitle

\begin{abstract}
Image demosaicing and super-resolution are two important tasks in color imaging pipeline. So far they have been mostly independently studied in the open literature of deep learning; little is known about the potential benefit of formulating a joint demosaicing and super-resolution (JDSR) problem. In this paper, we propose an end-to-end optimization solution to the JDSR problem and demonstrate its practical significance in computational imaging. Our technical contributions are mainly two-fold. On network design, we have developed a Residual-Dense Squeeze-and-Excitation Networks (RDSEN) supported by a pre-demosaicing network (PDNet) as the pre-processing step. We address the issue of \emph{spatio-spectral attention} for color-filter-array (CFA) data and discuss how to achieve better information flow by concatenating Residue-Dense Squeeze-and-Excitation Blocks (RDSEBs) for JDSR. Experimental results have shown that significant PSNR/SSIM gain can be achieved by RDSEN over previous network architectures including state-of-the-art RCAN. On perceptual optimization, we propose to leverage the latest ideas including \emph{relativistic discriminator} and pre-excitation perceptual loss function to further improve the visual quality of textured regions in reconstructed images. Our extensive experiment results have shown that Texture-enhanced Relativistic average Generative Adversarial Network (TRaGAN) can produce both subjectively more pleasant images and objectively lower perceptual distortion scores than standard GAN for JDSR. Finally, we have verified the benefit of JDSR to high-quality image reconstruction from real-world Bayer pattern data collected by NASA Mars Curiosity. 
\end{abstract}

\begin{IEEEkeywords}
Color imaging, Joint image demosaicing and super-resolution (JDSR), residual-dense squeeze-and-excitation network (RDSEN), perceptual optimization.
\end{IEEEkeywords}

%
\IEEEpeerreviewmaketitle

\section{Introduction}
\begin{figure}
    \centering
    \includegraphics[width=8cm]{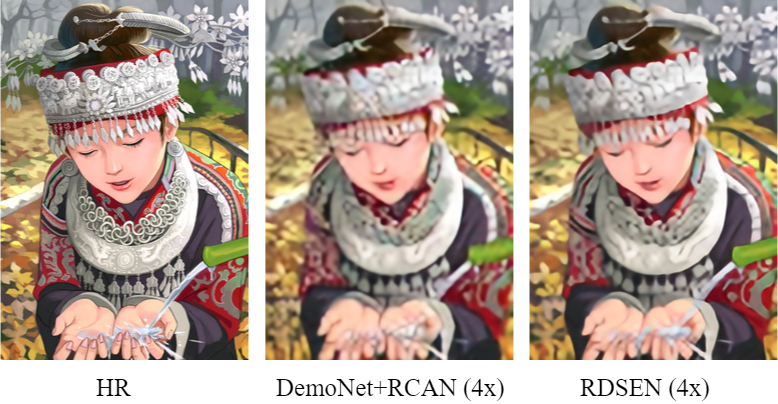}
    \caption{Comparison of JDSR output to separately demosaic-super-resovle output. Left to right: a) HR image (ground-truth); b) $4\times$ upscaling output by concatenating state-of-art demosaicing method DemoNet \cite{Gharbi_demos} with SISR method RCAN \cite{RCAN} (separated approach); c) $4\times$ upscaling output of our proposed RDSEN networks (joint approach).}
    \label{fig:intro}
\end{figure}

%
%
%
%
\IEEEPARstart{I}{mage} demosaicing and single image super-resolution (SISR) are two important image processing tasks to the pipeline of color imaging. Demosaicing is a necessary step to reconstruct full-resolution color images from so-called Color filter Array (CFA) such as Bayer pattern. SISR is a cost-effective alternative to more expensive hardware-based solution (i.e., optical zoom). Both problems have been extensively yet separately studied in the literature - from model-based methods \cite{zhang2005_demos, li_demos, li_demos2, ye2015_demos, bevilacqua2012low, chang2004super, timofte2013anchored, yang2010image, zeyde2010single} to learning-based approaches \cite{he2012self_demos, kapah2000_demos, Kokk_demos, sun2013_demos,VDSR,SRGAN,ESRGAN,RDN,RCAN}. Treating demosaicing and SISR as two independent problems may generate undesirable edge blurring as shown in Fig.~\ref{fig:intro}. Moreover, the processes of demosaicing and SISR can be integrated and optimized together from a practical application point of view (e.g., digital zoom for smartphone cameras such as iPhone 11 pro max, Google Pixel 4 and Huawei P30).

\begin{figure*}
    \centering
    \includegraphics[width=17cm]{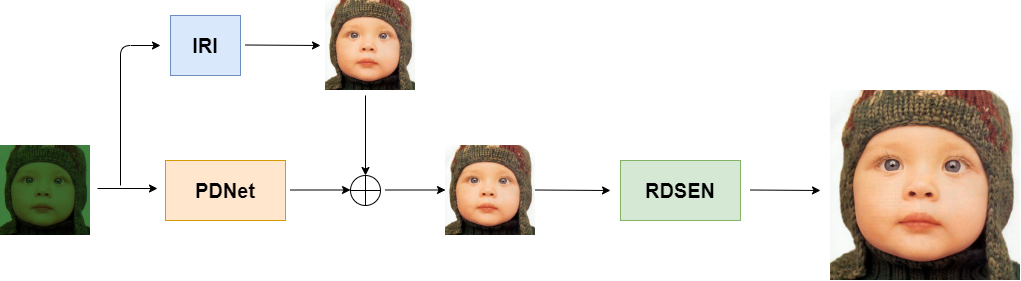}
    \caption{Overview of proposed RDSEN with PDNet network architecture, $\oplus$ means element-wise sum.}
    \label{fig:overview}
\end{figure*}

Inspired by the success of joint demosaicing and denoising \cite{Gharbi_demos}, we propose to study the problem of joint image demosaicing  and super-resolution (JDSR) in this paper and develop a principled solution leveraging latest advances in deep learning to computational imaging. We argue that the newly formulated JDSR problem has high practical impact (e.g., to support the mission of NASA Mars Curiosity and smartphone applications). The problem of JDSR is intellectually appealing but has been under-researched so far. The only existing work we can find in the open literature is a recently published paper \cite{RDSR} which contained a straightforward application of ResNet \cite{resnet} and considered the scaling ratio of two only. As demonstrated in Fig.~\ref{fig:intro}, our optimized solution to JDSR can achieve significantly better visual quality than the brute-force approach.  

The motivation behind our approach is mainly two-fold. On one hand, rapid advances in deep residual learning have offered a rich set of tools for image demosaicing and SISR. For example, DenseNet \cite{densenet} has been adapted to fully exploit hierarchical features for the problem of SR in SRDenseNet \cite{denseSR} and residual dense network (RDN) \cite{RDN}; residual channel attention network (RCAN) \cite{RCAN} allows us to develop much deeper networks (over 400 layers) with squeeze-and-excitation (SE) blocks \cite{SENet} than previous works (e.g., \cite{VDSR, EDSR}). Inspired by RDN and RCAN, our previous \cite{xu2019scan} presented a spatial color attention mechanism (SCAN) to further improve the SISR performance on real world SR dataset. However, to the best of our knowledge, the issue of \emph{spatio-spectral attention} mechanism has not been explicitly addressed for raw CFA data in the open literature. How to design a network architecture for \emph{jointly} exploiting spatial and spectral dependency in Bayer patterns deserves a systematic study. 

On the other hand, we propose to optimize the \emph{perceptual quality} for JDSR because that is what really matters in various real-world applications (e.g., to support the mission of NASA to Mars). Generative adversarial network (GAN) \cite{gan} is arguably the most popular approach toward perceptual optimization and has demonstrated convincing improvement for SISR in SRGAN \cite{SRGAN}. It has also been widely observed that the training of GAN suffers from stability issue which could have catastrophic impact on reconstructed images. There has been a flurry of latest works (e.g., Relativistic average GAN (RaGAN) \cite{RaGAN}, enhanced SRGAN (ESRGAN) \cite{ESRGAN} and perception-enhances SR (PESR) \cite{PESR}) showing the potential of relativistic discriminator in stabilizing GAN and improving visual quality of SISR images. However, the issue of perceptual optimization has not been addressed in previous works on joint demosaicing-and-denoising (JDD) at all \cite{Gharbi_demos}, \cite{kokkinos2018deep}, \cite{kokkinos2019iterative}. For the first time, we aim at studying the potential of GAN-based models on perceptual optimization for the JDSR problem, which has practical significance when no ground-truth (reference image) is available.

\begin{figure*}
    \centering
    \includegraphics[width=17cm]{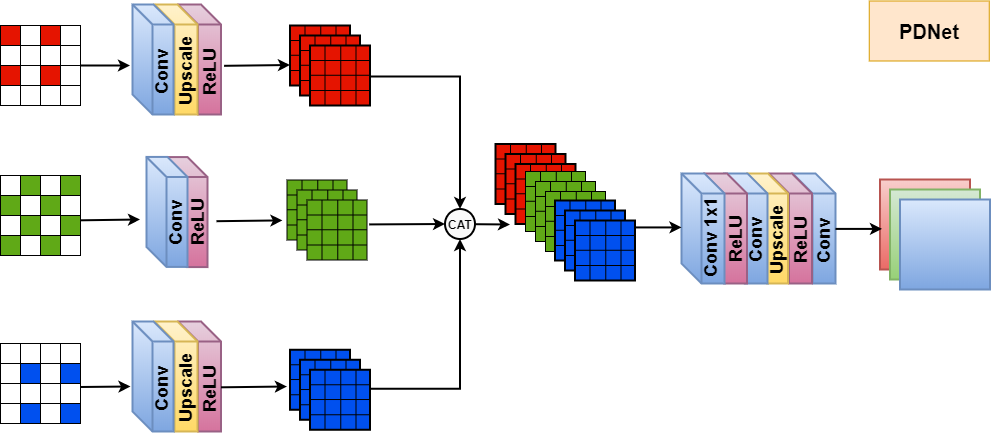}
    \caption{Structure of PDNet, `CAT' is feature concatenation.}
    \label{fig:PDNet}
\end{figure*}

Overall, our contributions are summarized as follows:

$\bullet$ Network design: we propose a concatenation of pre-demosaicing network (PDNet) and Residual-Dense Squeeze-and-Excitation Networks (RDSEN) for JDSR. The former takes a model-based demosaicing result via iterative-residual interpolation (IRI) \cite{ye2015color} as the surrogate target to facilitate deep residue learning for pre-demosaicing. Then a novel concatenation of Residual-Dense Squeeze-and-Excitation Block (RDSEB) modules is designed to facilitate information flow between the intermediate demosaicing result and the final reconstruction. Through the combination of long and short skip connections, we manage to train RDSEN more efficiently than existing RCAN while still achieving better performance. 

$\bullet$ Perceptual optimization: we have leveraged the latest advance RaGAN \cite{RaGAN} from SISR to JDSR and studied the choices of perceptual loss function for JDSR. In addition to improved stability, we have found that Texture-enhanced RaGAN (TRaGAN) with a before-activation perceptual loss function can produce visually more pleasant results. We argue that the issue of perceptual optimization is particularly important for JDSR because it has been largely overlooked in the existing literature of JDD.

$\bullet$ Simulation study and real-world application: We have conducted extensive stimulation study to demonstrate the superiority of our network to other competing approaches. When compared against the current state-of-the-art RCAN \cite{RCAN}, our RDSEN has achieved significant improvement on both objective (up to 1.2dB in terms of PSNR on $McM$ dataset) and subjective qualities. We have also applied the proposed RDSEN+TRaGAN solution to raw Bayer pattern data collected by the Mast Camera (Mastcam) of NASA Mars Curiosity Rover. Our experimental results have shown visually superior high-resolution image reconstruction can be achieved at the scaling ratio as large as 4.



 



\section{Related Works}
Both image demosaicing and super-resolution have been studied in decades in the open literature. In this section, we review image demosaicing and image super-resolution approaches separately and focus on deep learning based methods. 

\subsection{Image Demosaicing}

Existing approaches toward image demosaicing can be classified into two categories: model-based methods \cite{zhang2005_demos, li_demos, li_demos2, ye2015_demos} and learning-based methods  \cite{he2012self_demos, kapah2000_demos, sun2013_demos}. Model-based approaches rely on hand-crafted parametric models which often suffer from lacking of the generalization capability to handle varying characteristics in color images (i.e., the potential model-data mismatch). Recently, deep learning methods show the advantages in image demosaicing field. Inspired by single image super-resolution model SRCNN \cite{SRCNN}, DMCNN \cite{Syu_demos} utilized super-resolution based CNN model and ResNet \cite{resnet} to investigate image demosacing problem. CDM-CNN \cite{Tan_demos} introduced to apply residual learning \cite{resnet} with a two-phase network architecture which firstly recovers green channel as guidance prior and then uses this guidance prior to reconstruct the RGB channels. Besides to explore image demosacing methods only, there are several works studying joint image demosaicing and denoising (JDD) problem. Dong \emph{et.al.} \cite{dong2018joint} developed a deep neural network with generative adversarial networks (GAN) \cite{gan} and perceptual loss functions to solve JDD problems. Inspired by classical image regularization and majorization-minimization optimization,  Kokkinos and Lefkimmiatis \cite{Kokk_demos} proposed a deep neural network to solve JDD problem. Deep learning based image demosaicing techniques have shown convincingly improved performance over model-based ones on several widely-used benchmark dataset (e.g., Kodak and McMaster \cite{McM}). However, the issue of suppressing spatio-spectral aliasing has not been addressed in the open literature as far as we know.

\begin{figure*}
    \centering
    \includegraphics[width=17cm]{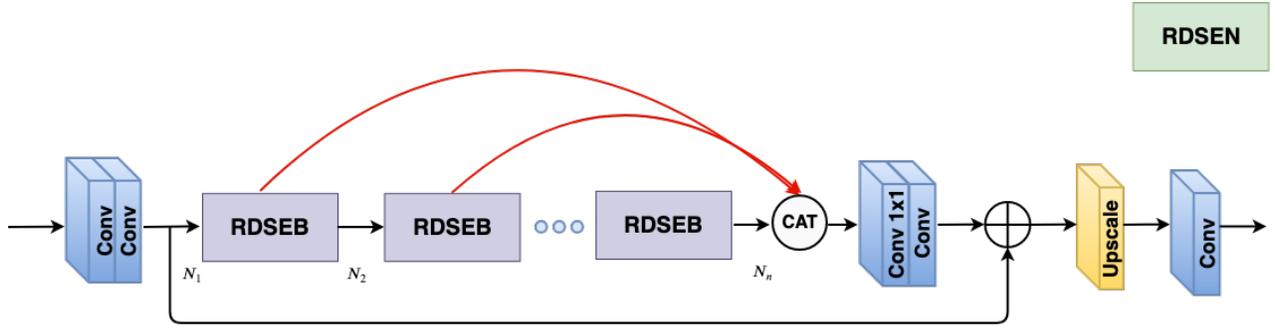}
    \caption{Structure of RDSEN, `CAT' is feature concatenation and $\oplus$ denotes element-wise sum respectively.}
    \label{fig:RDSEN}
\end{figure*}

\subsection{Image Super-resolution}

Model-based approaches towards SISR \cite{bevilacqua2012low, chang2004super, timofte2013anchored, yang2010image, zeyde2010single} suffer from notorious aliasing artifacts and edge blurring. Recently, deep learning-based approaches have advanced rapidly. SRCNN \cite{SRCNN} first introduced deep learning based method to solve single image super-solution task with three convolutional layers and achieved much better performance than model based methods. Benefit by concept of ResNet \cite{resnet}, VDSR \cite{VDSR} firstly trained 20 layers deep networks with long residual connection which can only learn more high-frequency information and increase the convergence speed. EDSR \cite{EDSR} proposed to integrate several resblocks and remove batch-normalization layer, which can save GPU memory, stack more layers and make networks wider \cite{singh2020going}, to further improve SISR performance. LapSRN \cite{laplace} proposed to super-resolve LR image several times to save GPU memory and achieve better performance. 

Most recent advances include SRDenseNet \cite{denseSR} which applied denseNet \cite{densenet} to solve SISR task, RDN \cite{RDN} which utilized ResNet and DenseNet to create residual dense block (RDB). Through local feature fusion, the proposed RDB can allow larger growth rate to boost the performance. RCAN \cite{RCAN} first introduced attention mechanism inspired by SENet \cite{SENet} to calibrate feature maps and proposed residual in residual structure to achieve a very deep convolutional networks which achieved new state-of-art performance for SISR task. Besides objective measures such as PSNR/SSIM \cite{ssim}, SRGAN \cite{SRGAN} introduced a novel generative adversarial networks (GAN) \cite{gan} based architecture to optimize the perceptual quality of SR images, benefit by GAN, SRGAN can reconstruct more textures from low-res images. An enhanced version of SRGAN named ESRGAN \cite{ESRGAN} using relativistic average GAN (RaGAN) was developed in \cite{RaGAN} as well as \cite{PESR} which can recover more realistic super-resolved image compared with SRGAN. By contrast, the problem of JDSR has been under-researched so far with the only exceptions of \cite{vandewalle2007joint}, \cite{RDSR}, and \cite{qian2019trinity}.


\section{Network Design: Residual-Dense Attention}
The hierarchy of our network design goes like: Overview of proposed network  (Fig. \ref{fig:overview}) $\rightarrow$ PDNet subnetwork (Fig.~\ref{fig:PDNet}) $\rightarrow$ RDSEN (Fig. \ref{fig:RDSEN}) $\rightarrow$ RDSEB with channel attention (Fig.~\ref{fig:RDSEB}).

\subsection{Pre-demosaicing Network}

One challenging issue in JDSR is that not only high-frequency components but also two-third of color pixels are missing. This issue can lead undesirable distortion or artifact in the reconstructed full-resolution color image. Inspired by recent work CBDNet \cite{guo2019toward}, we have designed a pre-demosaicing network (PDNet) for initially demosaicing the Bayer pattern as a pre-processing step to reduce the gap between LR CFA data and HR color image. As shown in Fig.~\ref{fig:overview} before the RDSEN module, we have adopted a model-based demosaicing method called iterative-residual interpolation (IRI) \cite{ye2015color} to generate an intermediate demosaicing result, which will be used as the input to the refinement module. This intermediate demosaicing results will be refined by PDNet as shown in Fig. \ref{fig:PDNet} (conceptually similar to ResNet \cite{resnet}). In the PDNet, we opt to separately process Red, Green, and Blue channels. For Red and Blue channel, we use a convolution layer with stride of 4 to shrink the corresponding Bayer pattern and then upscale them with a factor of 2; for Green channel we shrink it by a convolution layer with stride of 2. This is because the Red and Blue channels each contains one-fourth information and G channel contains one-half information. Then we concatenate  RGB feature maps and fused them with a $1 \times 1$ kernel of convolution layer. Finally we upscale the fused feature maps back to the same size as input CFA data.

\subsection{Residual-Dense Squeeze-and-Excitation Network}

Channel attention mechanism has been successfully applied to both high-level (e.g., SENet \cite{SENet} and LS-CNN \cite{ls-cnn}) and low-level (e.g., RCAN \cite{RCAN}) vision tasks. A channel attention module first squeezes the input feature map and then activates one-time reduction-and-expansion to excite the squeezed feature map. Such strategy is not optimal for recovering missing high-frequency information in SISR when the network is very deep; meanwhile, JDSR problem requires simultaneous recovery of incomplete color information across Red, Green, Blue channels, which requires extra attention toward the dependency in the spectral domain. How to generalize the channel attention mechanism from spatial-only to joint \emph{spatio-spectral} has remained one of open problems in the design of attention-based networks.

\begin{figure*}[t]
    \centering
    \includegraphics[width=17cm]{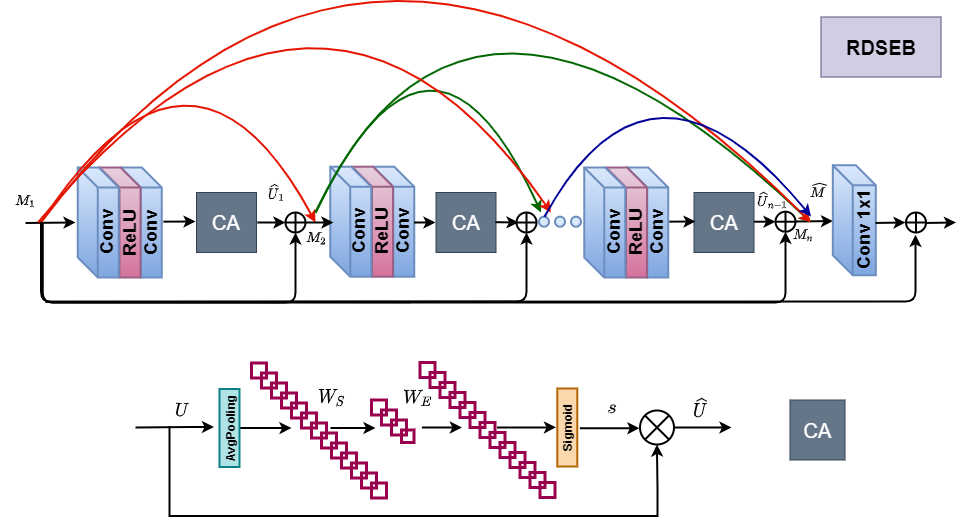}
    \caption{Flowchart of Residual-Dense Squeeze-and-Excitation Block (RDSEB) and Channel Attention (CA) module ($\otimes$ denotes element-wise product).}
    \label{fig:RDSEB}
\end{figure*}

As discussed in \cite{RCAN}, high-frequency components often correspond to regions in an image such as textures, edges, corners and so on. Conventional layers have limited capability of exploiting contextual information outside the {\em local} receptive field especially due to the missing data in Bayer patterns. To overcome this difficulty, we propose to design a new Residual-Dense Squeeze-and-Excitation Network (RDSEN) as shown in Fig.~\ref{fig:RDSEN} and Fig. ~\ref{fig:RDSEB}. The proposed RDSEN is designed to implement a \emph{deeper} and \emph{wider} spatio-spectral channel attention mechanism for the purpose of more effectively suppressing spatio-spectral aliasing in LR Bayer patterns. 

Unlike SENet \cite{SENet} and RCAN \cite{RCAN} (using residual block to stack with channel attention module), RDSEN based on multiple RDSEB blocks attemps to fuse both {\em local and global} residual-dense attention information to assure more faithful information recovery when the network gets deeper and wider. As shown in Figs.~\ref{fig:RDSEN} and~\ref{fig:RDSEB}, we have kept both long skip and short skip connections like RCAN in order to make the overall training stable and facilitate the information flow both inside and outside the RDSEN module. Although similar idea of local feature fusion existed in residual dense block of RDN \cite{RDN}, our {\em hybrid} design - i.e., the RDSEB block combining the ideas from RDN and RCAN - is novel because it represents an alternative approach to strike an improved tradeoff between cost (in terms of network parameters) and performance (in terms of visual quality). 

Our design of concatenating RDSEB modules also has its merit from the perspective of exploiting joint spatio-spectral attention for JDSR. Spatio-spectral channel attention mechanism in the proposed RDSEB module can help to recalibrate input features via channel statistics \cite{SENet} across different spectral bands. In SISR, residual-in-residual attention or dense connection operation might be sufficient for capturing channel-wise dependencies for LR color images; however our JDSR task aims at recovering two-third of missing data in spectral bands in addition to the missing high-frequency information. We have experimentally verified that such design of deeper and wider networks \cite{singh2020going} based on concatenation of multiple RDSEB modules indeed helps the boosting of our JDSR performance.

\subsection{Residual-Dense Squeeze-and-Excitation Block}

The key to deeper and wider networks lies in the design of RDSEB module - i.e., how to use short skip connection and multiple concatenations after channel attention mechanism to assure faithful information recovery both inside and outside RDSEB modules? As shown in Fig.~\ref{fig:RDSEB}, we propose a Residual-Dense Squeeze-and-Excitation Block (RDSEB)  in which the channel size can be expanded \emph{step by step} (see Fig.~\ref{fig:RDSEB}). The key advantages of this newly designed RDSEB include: 1) the reduced channel descriptor can be smoothly activated \emph{multiple times} and therefore more faithful information across spatio-spectral domain is accumulated; 2) dense-connection can increase the network \emph{depth and width} without running into the notorious vanishing-gradient problem \cite{bengio1994learning}; 3) both information flow and network stability, which are important to a principled solution to JDSR, can be jointly improved by introducing dense connections to SE residual blocks (so we can train even deeper than RCAN \cite{RCAN}).

More specifically, to implement the CA block, we first apply global average pooling to \emph{squeeze} input feature maps. Let us denote the input feature maps by $\textbf{U} = [u_1, u_2, ..., u_C]$, which contains $C$ feature maps with the dimension of $H\times W$. Then the global average pooling output $z \in \mathbb{R}^C $ can be calculated by:
\begin{equation}
    z_C = \frac{1}{H \times W}\sum_{i=1}^{H}\sum_{j=1}^{W}\textbf{u}_C(i,j)
\label{average pooling}
\end{equation}
where $z_C$ is the $c$-th element of $z$, $u_C(i,j)$ is the pixel value of the $c$-th feature at position $(i,j)$ from input feature maps. Then we propose to implement a simple gating mechanism as adopted by previous works including SENet \cite{SENet} and RCAN \cite{RCAN}:
\begin{equation}
\label{eqn:CA}
s = \sigma(\textbf{W}_E(\delta(\textbf{W}_S(z))))
\end{equation}
where $\sigma$ refers to a sigmoid function, $\delta$ denotes the ReLU function, both $\textbf{W}_S$ and $\textbf{W}_E$ are Conv layers with weights $\textbf{W}_S \in \mathbb{R}^{1 \times 1 \times \frac{C}{r}}$ and $\textbf{W}_E \in \mathbb{R}^{1 \times 1 \times C}$, $r$ is the scaling ratio to reduce the dimension of $z$ (details about this hyperparameter controlling the tradeoff between the capacity and the complexity can be found in SENet \cite{SENet}).

In order to achieve deeper and wider channel attention, we propose a novel strategy of connecting each output of channel attention block not only with short skip connection (residual) but also dense-connection as shown in Fig.~\ref{fig:RDSEB}. To formalize this problem, define $\text{U}$ as the input feature map to CA module, the rescaled input feature map $\hat{\textbf{U}}$ can be expressed as: 

\begin{equation}
\hat{\textbf{U}} = s \cdot \text{U}
\end{equation}
where `$\cdot$' stands for element-wise product.

Finally, to implement dense-connection, we define $M_{1}$ as the input feature map of RDSEB block. Then the output feature map $\hat{M}$ can be written as the following equations: 

\begin{alignat}{2}
   & M_{i} =\hat{\textbf{U}}_{i-1} + M_{1}, \text{ where } i \in [2,n] \\
   & \hat{M} = [M_{1}, M_{2},..., M_{n}]
\end{alignat}
where $[M_{1}, M_{2}...M_{n}]$ refers to the concatenation of feature maps, $\hat{\textbf{U}}_{i-1}$ is the corresponding output of CA module at the $i-1$-th stage as shown in Fig.~\ref{fig:RDSEB}. With the new RDSEB block, we can train a deeper and wider network thanks to the improved information flow.


\begin{table*}
\small
\begin{center}
\resizebox{\textwidth}{!}{
\begin{tabular}{l|c|c|c|c|c|c|c}
\hline
\multirow{2}{*}{Method} &   \multirow{2}{*}{Scale}  &  Set5  & Set14 & B100  & Manga109 & McM & PhotoCD\\
\cline{3-8} & & PSNR/SSIM & PSNR/SSIM  & PSNR/SSIM & PSNR/SSIM & PSNR/SSIM & PSNR/SSIM \\
\hline\hline
FlexIPS\cite{FlexISP}+RCAN\cite{RCAN}  &  x2  &  35.18/0.9387 &  31.24/0.8776  & 31.00/0.8647    &   30.32/0.9199 &  34.80/0.9301  &  43.02/0.9610  \\
DemoNet\cite{Gharbi_demos}+RCAN\cite{RCAN} & x2  &   35.92/0.9458   &  32.27/0.8971   & 31.38/0.8823   &  35.50/0.9590 & 35.34/0.9362   &  43.53/0.9642       \\
RDSR\cite{RDSR}                        & x2   & 36.29/0.9485   &  32.56/0.9008  & 31.56/0.8850   & 36.14/0.9625  &  35.90/0.9423 &  43.74/0.9655    \\
RCAN \cite{RCAN}                                           & x2  & \underline{36.54}/\underline{0.9499}   &   \underline{32.74}/\underline{0.9032}  &  \underline{31.68}/\underline{0.8878}  & \underline{36.65}/\underline{0.9643}   & \underline{36.18}/\underline{0.9445}  &   \underline{43.91}/\underline{0.9661}   \\
RDSEN (ours) & x2 & \textbf{37.40/0.9575} &\textbf{32.91/0.9128} & \textbf{32.00/0.8972}&  \textbf{36.86/0.9716} &\textbf{37.38/0.9565} &\textbf{44.70/0.9716}  \\
\hline\hline
FlexISP+RCAN & x3 & 31.21/0.8731 & 28.55/0.7884 & 27.31/0.7310  & 27.58/0.8647 & 31.25/0.8661 & 40.32/0.9402 \\
DemoNet+RCAN & x3  & 32.16/0.9030    &  29.24/0.8137   & 28.42/0.7801   & 30.75/0.9112   & 31.65/0.8739   & 40.74/0.9445    \\
RDSR                                         & x3 &  33.05/0.9103 &  29.54/0.8211  &   28.61/0.7859 &    31.69/0.9225 & 32.21/0.8842  & 40.90/0.9458     \\
RCAN                                        & x3 &  \underline{33.24}/\underline{0.9125}  &  \underline{29.67}/\underline{0.8241}  &  \underline{28.69}/\underline{0.7882}   & \underline{32.06}/\underline{0.9267}  & \underline{32.42}/\underline{0.8874}  &  \underline{41.11}/\underline{0.9469}    \\
RDSEN (ours) & x3 & \textbf{33.75/0.9218} & \textbf{29.91/0.8337}  & \textbf{28.84/0.7993} &  \textbf{32.14/0.9330} & \textbf{33.21/0.9032 }& \textbf{41.60/0.9521} \\
\hline\hline
FlexISP+RCAN  & x4 &  29.57/0.8376  &  26.94/0.7177  &  26.68/0.6896   &    26.69/0.8427  & 27.78/0.7651   & 38.28/0.9201    \\
DemoNet+RCAN & x4  &  30.33/0.8596    & 27.58/0.7488   &   26.94/0.7081 &  27.81/0.8590& 29.49/0.8187   &    38.67/0.9243    \\
RDSR                                       & x4  &30.87/0.8712   &27.91/0.7589    &27.16/0.7151    &   28.86/0.8800  & 30.10/0.8328  &  38.81/0.9258    \\
RCAN                                      & x4 &\underline{31.04}/\underline{0.8746}   &  \underline{27.98}/\underline{0.7613} &  \underline{27.20}/\underline{0.7175}  &   \underline{29.12}/\underline{0.8856}  & \underline{30.24}/\underline{0.8367}  &   \underline{39.01}/\underline{0.9271}  \\
RDSEN (ours) & x4 & \textbf{31.63/0.8863} & \textbf{28.26/0.7725} & \textbf{27.39/0.7284} &  \textbf{29.28/0.8903} & \textbf{30.74/0.8523} & \textbf{39.41/0.9317} \\

\hline
\end{tabular}
}
\end{center}
\caption{PSNR/SSIM comparison among different competing methods. \textbf{Bold} font indicates the best result and \underline{underline} the second best.}
\label{tab:PSNR}
\end{table*}

\section{Perceptual Optimization: Relativistic Discriminator and Loss Function}

\subsection{Texture-enhanced Relativistic average GAN (TRaGAN)}

The discriminator $D$ in standard GAN \cite{gan} only estimates the probabilities of real/fake images, and the interaction between generator and discriminator is interpreted as a two-player minimax game. It can be expressed as $D(x) = \sigma(C(x))$, where $\sigma$ is sigmoid function, $C(x)$ is non-transformed layer, $x$ is the input image. Such idea has been successfully applied to the problem of SISR such as SRGAN \cite{SRGAN} in which the super-resolved image (fake version) is compared against the ground-truth (real version). In other words, discriminator $D$ serves as a judge for perceptual optimization of generator.

Unlike standard GAN, relativistic average GAN (RaGAN) \cite{RaGAN} can make the discriminator $D$ to estimate the probability based on both real and fake images, making a real image more realistic than a fake one (on the average). According to \cite{RaGAN}, RaGAN can not only generate more realistic images but also stabilize the training progress. Recently, the benefit of RaGAN over conventional GAN has been demonstrated for SISR in \cite{ESRGAN} and \cite{PESR}. Here we propose to leverage the idea of RaGAN to  JDSR and demonstrate how relativistic discriminator can work with the proposed RDSEN (generator) for the purpose of perceptual optimization when no ground-truth (reference image) is available. Note that this issue has been overlooked in the literature of not only SISR (e.g, RDN \cite{RDN}, RCAN \cite{RCAN}) but also JDD (e.g., \cite{Gharbi_demos}, \cite{kokkinos2018deep}, \cite{kokkinos2019iterative}).

To implement RaGAN, we represent the real and fake images by $x_r$ and $x_f$ respectively; then we can formulate the output of a modified discriminator $\hat{D}$ for RaGAN by:
\begin{alignat}{2}
\label{real_fake}
& \hat{D}(x_r) = \sigma(C(x_r) -\mathbb{E}_{x_f}[C(x_f)]) \\
& \hat{D}(x_f) = \sigma(C(x_f) - \mathbb{E}_{x_r}[C(x_r)])
\end{alignat}
where $\mathbb{E}_{x_f}$ and $\mathbb{E}_{x_r}$ are the expectation functions. It follows that the discriminator loss function $L_D^{RaGAN}$ and adversarial loss function $L_G^{RaGAN}$ can be written as:
\begin{alignat}{2}
&L_D^{RaGAN} = - \mathbb{E}_{x_r}[\log(\hat{D}(x_r)]- \mathbb{E}_{x_f}[\log(1-\hat{D}(x_f))] \\
&L_G^{RaGAN} = -\mathbb{E}_{x_r}[\log(1-\hat{D}(x_r))]- \mathbb{E}_{x_f}[\log(\hat{D}(x_f)]
\label{RaGAN_loss}
\end{alignat}

It has been observed that the class of texture images is often more difficult for SISR due to spatial aliasing \cite{PESR}. One way of achieving better texture reconstruction is through attention mechanism at the image level - i.e., to emphasize (i.e., increase the weight) difficult samples and overlook (i.e., down-weighting) easy ones. Such idea of weighting can be conveniently incorporated into the RaGAN package because the PyTorch implementation allows an optional weight input. More specifically, we propose to consider the following weighted function with a new hyperparameter $\gamma$ tailored for \emph{T}exture enhancement:
\begin{equation}
\begin{aligned}
L_G^{TRaGAN} = & -\sum_{i}(\hat{D}(x_r))^{\gamma}\log(1-\hat{D}(x_r)) \\
& - \sum_{i}(1-\hat{D}(x_f))^{\gamma}\log(\hat{D}(x_f)) 
\end{aligned}
\label{TRaGAN_loss}
\end{equation}

\subsection{Perceptual Loss Function}

We have implemented the following perceptual loss function based on \cite{perceptual, SRGAN, ESRGAN, PESR}.  With a pre-trained VGG19 model \cite{vgg}, we can extract high-level perceptual features of both high-resolution (HR) and SR images from the 4-$th$ convolutional layer of VGG19 before the activation function is applied. Inspired by \cite{ESRGAN}, we propose to extract high-level features before the activation function layer because it can further improve the performance. Let's define perceptual loss as $L_{vgg}$ and $L_1$-norm distance as $L_1$. Then the total loss for our generator $L_G$ can be formulated as follows:
\begin{equation}
L_G = L_{vgg} + \lambda_1  L_G^{TRaGAN} + \lambda_2 L_1
\label{total_loss}
\end{equation}
where coefficients $\lambda_1$ and $\lambda_2$ are used to balance different loss terms. $L_{vgg} = \Phi(f(SR), f(HR))$. $\Phi$ denotes the mean-squared error function (MSE), $f(SR)$ and $f(HR)$ are the high-level features extracted from the output of the $ 4^{th}$ convolution layer of VGGNet before the pooling. Note that although similar loss functions were considered in previous studies including \cite{ESRGAN} and \cite{PESR}, their experiments include synthetic low-resolution images only. In this paper, we will demonstrate the effectiveness of the proposed perceptual optimization for JDSR on both synthetic and real-world data next.

\begin{table*}
\small
\begin{center}
\begin{tabular}{l|c|c|c|c|c|c|c}
\hline
Methods &  Scale  &   Set5  & Set14 & B100  & Manga109 & McM & PhotoCD\\
\hline\hline
FlexISP\cite{FlexISP}+RCAN\cite{RCAN}  &  x2  &  4.16 &  4.14 &  3.34    &  4.97  &  3.51 & 5.42  \\
DemoNet\cite{Gharbi_demos}+RCAN\cite{RCAN} & x2  &  4.13   & 3.76   & 3.31    & 3.99  & 3.48  &   5.59     \\
RDSEN  (ours)        & x2 &  4.17 & 3.81  &   3.28  &  4.07  & 3.27  &  5.65 \\
RDSEN\_GAN  (ours)          & x2 &  \underline{3.41}  &  \underline{2.95}   &  \textbf{2.34}   &  \underline{3.53}  &  \underline{2.59} &  \underline{4.85} \\
RDSEN\_TRaGAN  (ours)          & x2 &  \textbf{3.06} & \textbf{2.90}  &  \underline{2.35}   & \textbf{3.45}   & \textbf{2.52}  &  \textbf{4.72} \\
\hline\hline
FlexISP\cite{FlexISP}+RCAN\cite{RCAN} & x3 & 6.98 & 5.70 & 6.18  & 5.43 & 5.14 & 6.42 \\
DemoNet+RCAN & x3  &  6.31   &  5.18 & 4.97  &4.63  & 5.19   &   6.61    \\
RDSEN  (ours) & x3 & 5.71 &4.74  & 4.48  & 4.53 & 4.57  & 6.52 \\
RDSEN\_GAN (ours) & x3 & \underline{3.78} & \underline{2.94} & \underline{2.39}   & \underline{3.44} & \underline{2.60} & \underline{4.96}\\
RDSEN\_TRaGAN (ours) & x3 & \textbf{3.58} & \textbf{2.81}  & \textbf{2.36}  & \textbf{3.37} & \textbf{2.44} & \textbf{4.78} \\
\hline\hline
FlexISP\cite{FlexISP}+RCAN\cite{RCAN}  & x4 & 7.42  &  6.63 &  6.30  &  5.28  &  7.15 &  6.88  \\
DemoNet+RCAN & x4  & 7.21    &  6.23  & 6.28  & 5.43 & 6.22  &   7.04    \\
RDSEN  (ours)      & x4  & 6.18 & 5.94 &5.92   & 5.00 & 5.68 &  6.87 \\
RDSEN\_GAN (ours)  & x4 & \underline{4.50} & \underline{3.31}& \underline{2.84} & \underline{3.65} & \underline{2.84} &  \underline{5.01}  \\
RDSEN\_TRaGAN (ours)  & x4 & \textbf{4.24} &  \textbf{3.11} &\textbf{2.55}  &  \textbf{3.45} & \textbf{2.72} & \textbf{4.44}  \\
\hline
\end{tabular}
\end{center}
\caption{Objective performance comparison among different methods in terms of Perceptual Index (the lower the better). \textbf{Bold} indicates the best result and \underline{underline} the second best.}
\label{tab:PI results}
\end{table*}

\section{Experimental results}

\subsection{Implementation details}

In our proposed RDSEN networks, we set the number of RDSEB blocks as 16; and each block includes 6 residual-dense SE modules. Most kernel size of Conv layers is $ 3 \times 3 $ with 64 filters ($ C = 64 $) except those described in particular: the Conv layers in CA modules and Conv layers marked as `$1\times1$' with a $1\times1$ kernel size. The reduction ratio is $r = 16$. The upscale module we have used is the same as \cite{ESPCNN}. The last layer filter is set to 3 in order to output super-resolved color images. For the discriminator setting, we have implemented the same discriminator network structure as SRGAN \cite{SRGAN}. All kernel size of Conv layers is $3\times3$.

In our PyTorch implementation of RDSEN, we first randomly crop the Bayer patterns as small patches with the size of 48 $\times$ 48, and crop the corresponding HR color images, with a batch size of 16; then we augment the training set by standard geometric transformations (flipping and rotation). Our model is trained and optimized by ADAM \cite{adam} with $ \beta_1 = 0.9 $, $ \beta_2 = 0.999 $, and $ \epsilon = 10^{-8} $. The initial learning rate is set to \num{1e-4}, the decay factor is set to 5, which decreases the learning rate by half after  [$ 80k $, $120k $, $150k$, $180k$] steps; the $L_1$ loss function is applied to minimize the error between HR and SR images. To train GAN-based networks, we have used the trained RDSEN to initialize the generator of GAN to get a better initial SR image for discriminator. The same learning rate and decay strategies are adopted here. $\lambda_1$ and $\lambda_2$ in Eq.~\eqref{total_loss} are set to $5\times 10^{-3}$ and $1\times 10^{-2}$ respectively as \cite{ESRGAN}.

Because the codes of RDSR \cite{RDSR} are not publicly available, we have tried our best to reproduce RDSR using PyTorch while keeping the batch size (16), patch size ($64\times 64$) and the number of residual blocks (24) exactly the same as used by the original work \cite{RDSR}. The learning rate and decay steps in RDSR implementation are the same as those in our RDSEN. This way, we have striven to make the experimental comparison against RDSR \cite{RDSR} as fair as possible.


\subsection{Training Dataset}  \label{section 5.2}

In our experiment, we have used DIV2K dataset \cite{div2k} as the training set, which includes 800 images (2K resolution). For testing, we have evaluated both popular image super-resolution benchmark datasets including Set5 \cite{bevilacqua2012low}, Set14 \cite{set14}, B100 \cite{b100}, and Manga109 \cite{manga109}, and popular image demosaicing datasets such as McMaster \cite{McM} and Kodak PhotoCD. To pre-process training and testing data, we downsample original high-resolution images by a factor of $2\times$, $3\times$, $4\times$ using Bicubic interpolation then generate the `RGGB' Bayer pattern. Based on previous work \cite{Syu_demos} and our own study (refer to next paragraph), supplying three-channels separately as the input (instead of the mosaicked single-channel composition) works better for the proposed network architecture. All experiments are implemented using PyTorch framework \cite{pytorch} and trained on NVIDIA Titan Xp GPUs. As an indicator of the overall computational complexity, the training time of our RDSEN lies somewhere between that of RDN \cite{RDN} and RCAN \cite{RCAN} as shown in Table.~\ref{tab:compare}. We have verified for all competing networks, it takes around 1000 epochs to reach the convergence.

\begin{table}
\small
    \centering
    \begin{tabular}{|c|c|c|c|}
    \hline
        Method &  RDN & RCAN &  RDSEN \\
        \hline
        Time (s) & 128 & 160  &  130 \\
        \hline
    \end{tabular}
    \caption{Training time compassion of RCAN, RDN and proposed RDSEN, per epoch }
    \label{tab:compare}
\end{table}


\begin{figure}[!t]
    \centering
    \includegraphics[width=8cm]{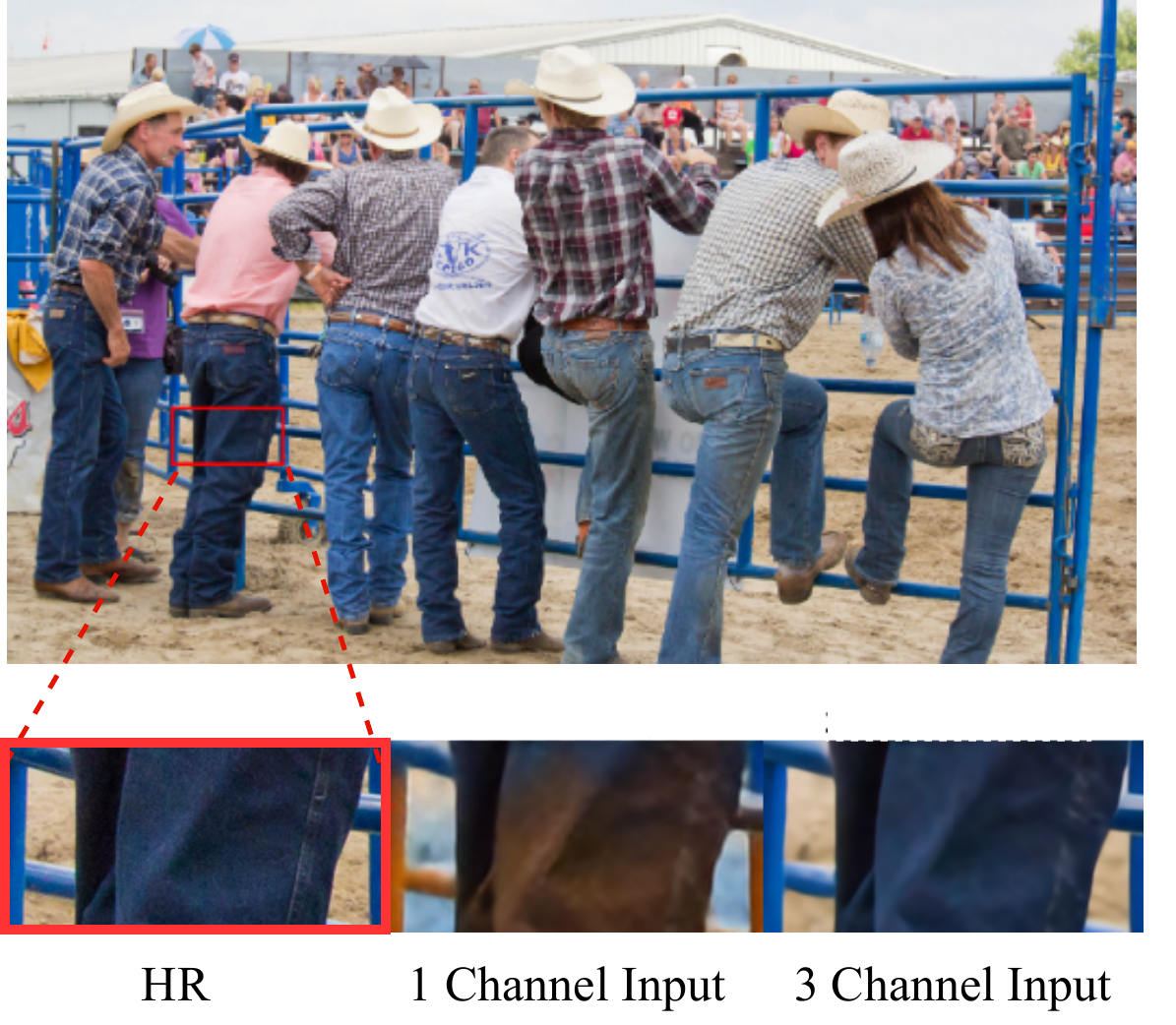}
    \caption{Visual comparison of training data effect, the bottom images, from left to right, are HR image, SR image generated by one-channel feature map (raw Bayer-pattern), SR image generated by three-channel feature map (R,G,B with zero padding for the missing pixels).}
    \label{fig:channel}
\end{figure}

\begin{figure*}
    \centering
    \includegraphics[width=17.5cm]{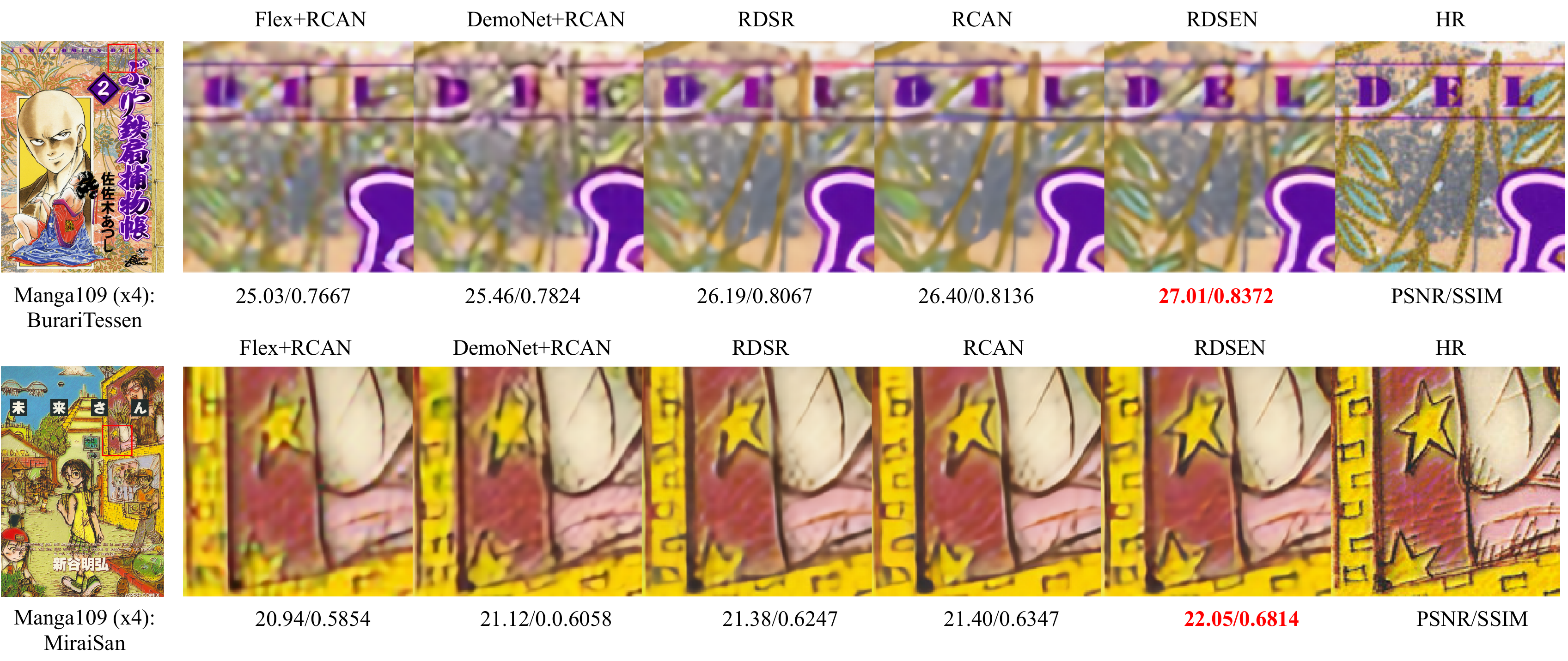}
    \caption{Visual results among competing approaches for Manga109 dataset at a scaling factor of 4.}
    \label{fig:PSNR2}
\end{figure*}

\begin{figure*}
    \centering
    \includegraphics[width=17.5cm]{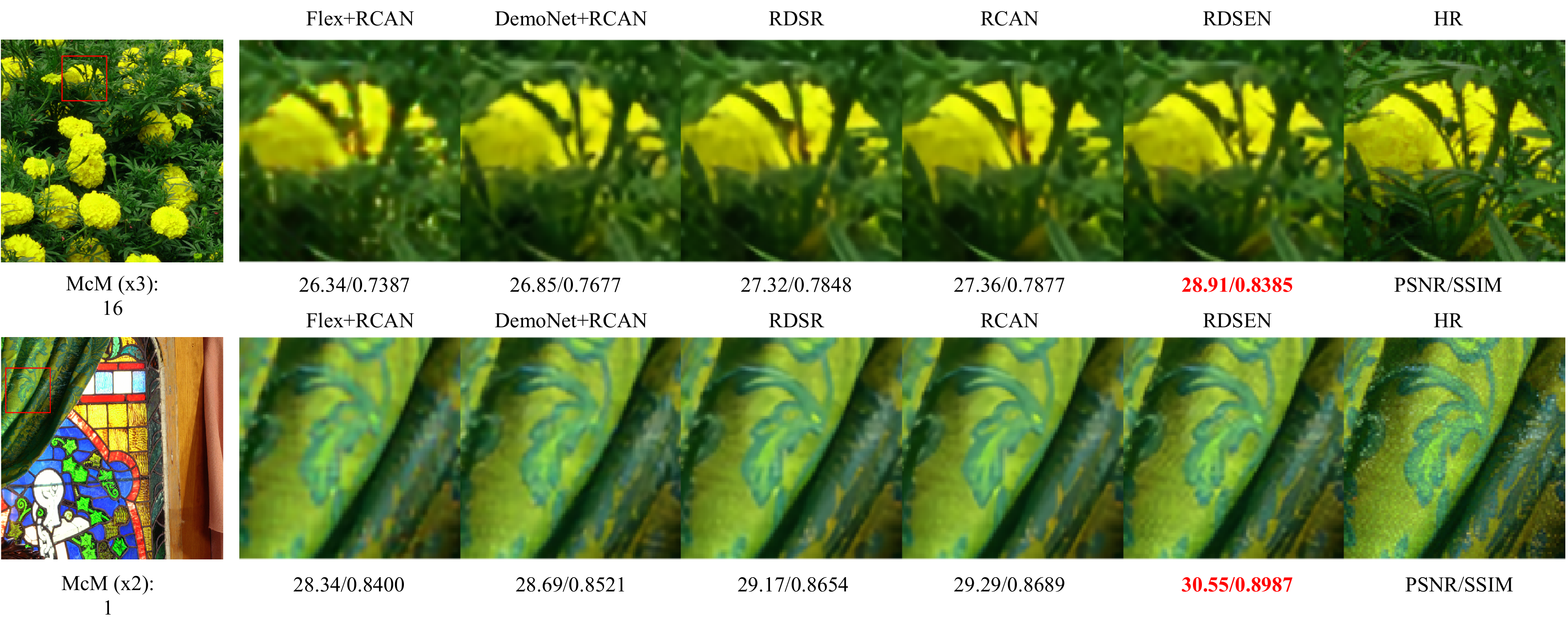}
    \caption{Visual results among competing approaches for McM atasets at a scaling factor of 2 and 3.}
    \label{fig:PSNR1}
\end{figure*}

\begin{figure*}
    \centering
    \includegraphics[width=17.5cm]{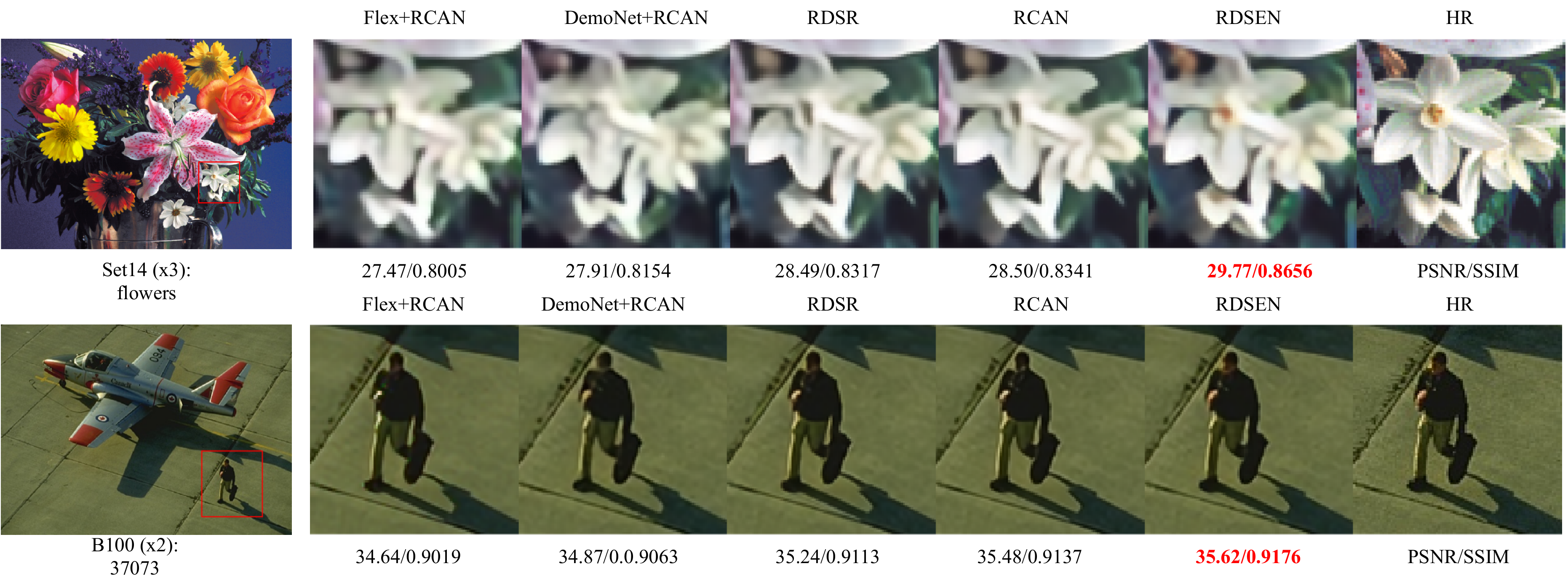}
    \caption{Visual results among competing approaches for Set14 and B100 datasets at a scaling factor of 3 and 2.}
    \label{fig:PSNR3}
\end{figure*}

Note that we have to be careful about four different spatial arrangements of Bayer patterns \cite{gunturk2005demosaicking}) in our definition of feature maps. One can either treat the Bayer pattern like a gray-scale image (one-channel setting) which ignores the important spatial arrangement of R/G/B; or take spatial arrangement as a priori knowledge and pad missing values across R,G,B bands by zeroes (three-channel setting). As shown in Fig. \ref{fig:channel}, the former has the tendency of producing color misregistration artifacts, which suggests the latter works better. Our experimental result has confirmed a similar finding previously reported in \cite{Syu_demos}. 
\subsection{PSNR/SSIM Comparisons}

\begin{figure*}
    \centering
    \includegraphics[width=16cm]{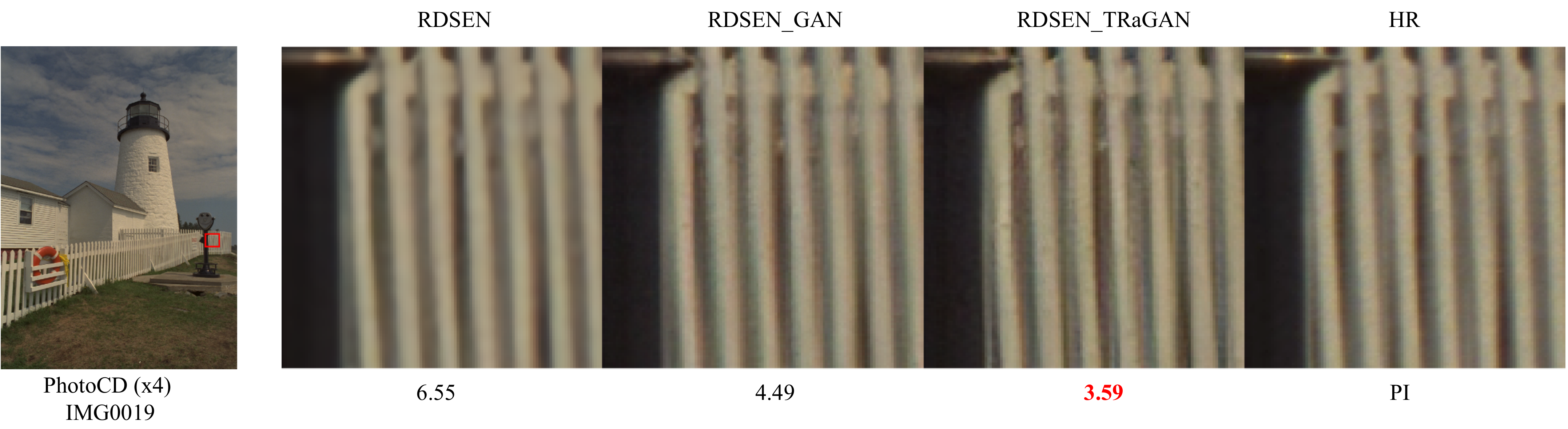}
    \caption{Visual comparison results among competing approaches for PhotoCD dataset at a scaling factor of 4.}
    \label{fig:GAN}
\end{figure*}

\begin{figure*}
    \centering
    \includegraphics[width=17cm]{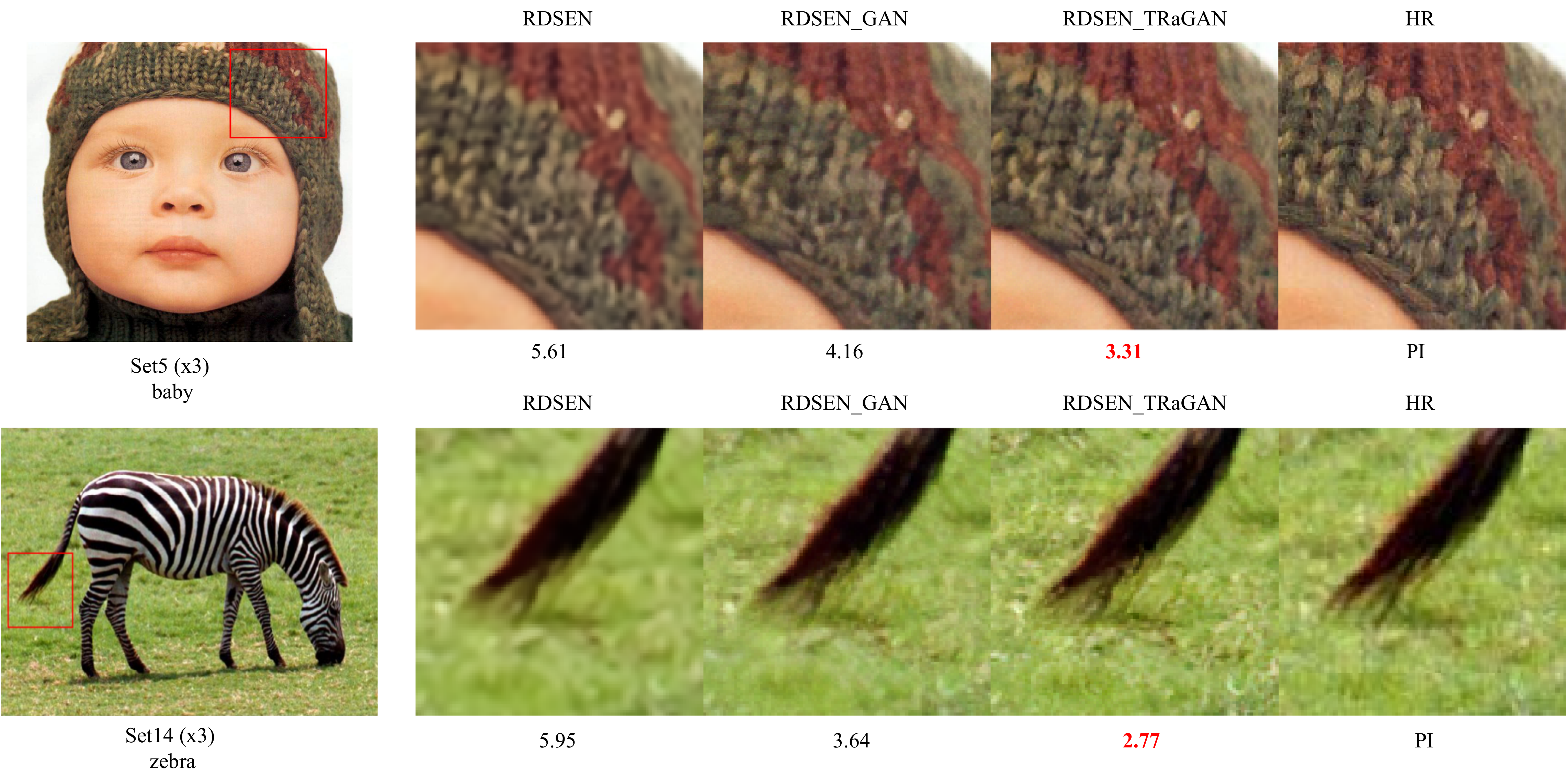}
    \caption{Visual comparison results among competing approaches for Set5 and Set14 datasets at a scaling factor of 3.}
    \label{fig:GAN2}
\end{figure*}
We have compared our methods against four benchmark methods: separated (brute-force) approaches Flex \cite{FlexISP} + RCAN \cite{RCAN} and DemoNet \cite{Gharbi_demos} + RCAN \cite{RCAN}, recently published literature RDSR \cite{RDSR}, and state-of-the-art SR approach RCAN \cite{RCAN}. 
To evaluate the results of DemoNet \cite{Gharbi_demos} + RCAN \cite{RCAN} approach, we first demosaiced the LR mosaiced images by using a pre-trained demosaicing network DemoNet to get LR color images, then super-resolved them by applying a pre-trained RCAN model. Note that we have used the pre-trained DemoNet and RCAN weights provided by the authors on GitHub. 



Table~\ref{tab:PSNR} shows PSNR/SSIM comparison results for scaling factors of $2\times$, $3\times$ and $4\times$.  It can be seen that our RDSEN method perform the best for all datasets and scale factors. We observe that significant PSNR/SSIM gains (up to $1.2dB$) over previous state-of-the-art. Since PSNR/SSIM metrics do not always faithfully reflect the visual quality of images, we have also included the subjective quality comparison results for images ``BurariTessen" and ``MiraiSan" in Fig.~\ref{fig:PSNR2}. For the first row of Fig.~\ref{fig:PSNR2}, it can be readily observed that for the top of the letters, only our RDSEN can faithfully recover text details; brute-force approaches (Flex+RCAN and DemoNet+RCAN), RDSR and RCAN have produced severe blurring artifacts; for the second row, only our method can reconstruct the yellow stars faithfully. Taking another example, Fig.~\ref{fig:PSNR1} shows the comparison at two other scaling factors ($3\times$ and $2\times$). For ``McM(x3)\_16", we observe that all approaches contain color artifacts between the flower and grass, but our RDSEN method can recover more realistic details than other competing approaches; for ``McM(x2)\_1'', pattern recovered by RDSEN appears to have the highest quality and most detailed textures. For more visual comparison, see Fig.~\ref{fig:PSNR3} which shows more convincing visual comparison among various competing approaches (please zoom in for detailed evaluation).

\subsection{Perceptual Index (PI) Comparisons}
Most recently, a new objective metric called Perceptual Index (PI) \cite{PI} has been developed for perceptual SISR (e.g., the 2018 PIRM Challenge \cite{blau:ECCVW:2018}). The PI score is defined by
\begin{equation}
\text{PI} = \frac{1}{2}((10-\text{MA})+\text{NIQE})
\end{equation}
where MA denotes a no-reference quality metric \cite{maScore} and NIQE referred to Natural Image Quality Evaluator \cite{NIQE}. Note that the lower PI score, the better perceptual quality (i.e., contrary to SSIM metric \cite{ssim}). Objective comparison of competing JDSR methods in terms of PI is shown in Table~\ref{tab:PI results}. We have observed that GAN-based methods produce the lowest PI scores for all datasets and scaling factors. Fig.~\ref{fig:GAN} provides the visual comparison with image "IMG0019" ($4\times$). It can be observed that GAN-based methods can recover sharper edges and overcome the issue of over-smoothed regions. Additionally, TRaGAN is capable of achieving even lower PI scores than the standard GAN. Fig.~\ref{fig:GAN2} shows another two results to demonstrate the advanced ability to recover texture details of GAN based methods, especially of TRaGAN.

\begin{table*}
\small
\begin{center}
\begin{tabular}{l|c|c|c|c|c|c}
\hline
\multirow{2}{*}{Method} &   \multirow{2}{*}{Scale}  &  Set5  & Set14 & B100  & Manga109 & McM \\
\cline{3-7} & & PSNR/SSIM & PSNR/SSIM & PSNR/SSIM & PSNR/SSIM & PSNR/SSIM\\
\hline\hline
ResNet  &  x2  &  36.48/0.9498 &  32.71/0.9030  & 31.67/0.8876    &   36.48/0.9642 &  36.11/0.9443  \\

RCAN                                           & x2  & 36.54/0.9499   &  32.74/0.9032  & 31.68/0.8878  &36.65/0.9643   & 36.18/0.9445   \\

RDSEN   (ours)            &  x2  &  \textbf{37.40}/\textbf{0.9575} &  \textbf{32.91}/\textbf{0.9128}  & \textbf{32.00}/\textbf{0.8972}   &   \textbf{36.86}/\textbf{0.9716} & \textbf{37.38}/\textbf{0.9565}  \\
\hline
\end{tabular}
\end{center}
\caption{Ablation study for ResNet, ResNet with CA (RCAN) and ResNet with proposed RDSEN. \textbf{Bold} font indicates the best result.}
\label{tab:ab}
\end{table*}

\subsection{Ablation Studies}

To demonstrate the effect of proposed RDSEB module, we study the networks: 1) only based on ResNet; 2) ResNet with channel attention module (RCAN); 3) ResNet with proposed Dense connected Squeeze-and-Excitation modules (RDSEN). All three networks are trained under same setting for fair comparison. The general SR benchmark datasets are used, scale factor is 2. From Table.~\ref{tab:ab}, we have found that ResNet has similar performance to more advanced RCAN. But when compared with our proposed RDSEN, the PSNR/SSIM performance of RCAN and ResNet are much lower than RDSEN; the proposed RDSEN has the best performance on all benchmark datasets.

\subsection{Performance on the Real-world Data}
Finally, we have tested our proposed JDSR technique on some real-world data collected by the Mastcam of NASA Mars Curiosity. The raw data are `RGGB' bayer pattern sized by $1600\times1200$. Due to hardware constraints, the left camera and the right camera of Mastcam have different focal lengths (the left is about 3 times weaker than the right). To compensate such a ``lazy-eye'' effect on raw Bayer patterns, it is desirable to develop a joint demosaicking and SR technique with at least a scaling factor of 3 (in order to support high-level standard stereo-based vision tasks such as 3D reconstruction and object recognition). Our proposed JDSR algorithm is a perfect fit for this task, which shows the great potential of computer vision and deep learning in deep space exploration. 

\begin{figure*}
    \centering
    \includegraphics[width=17cm]{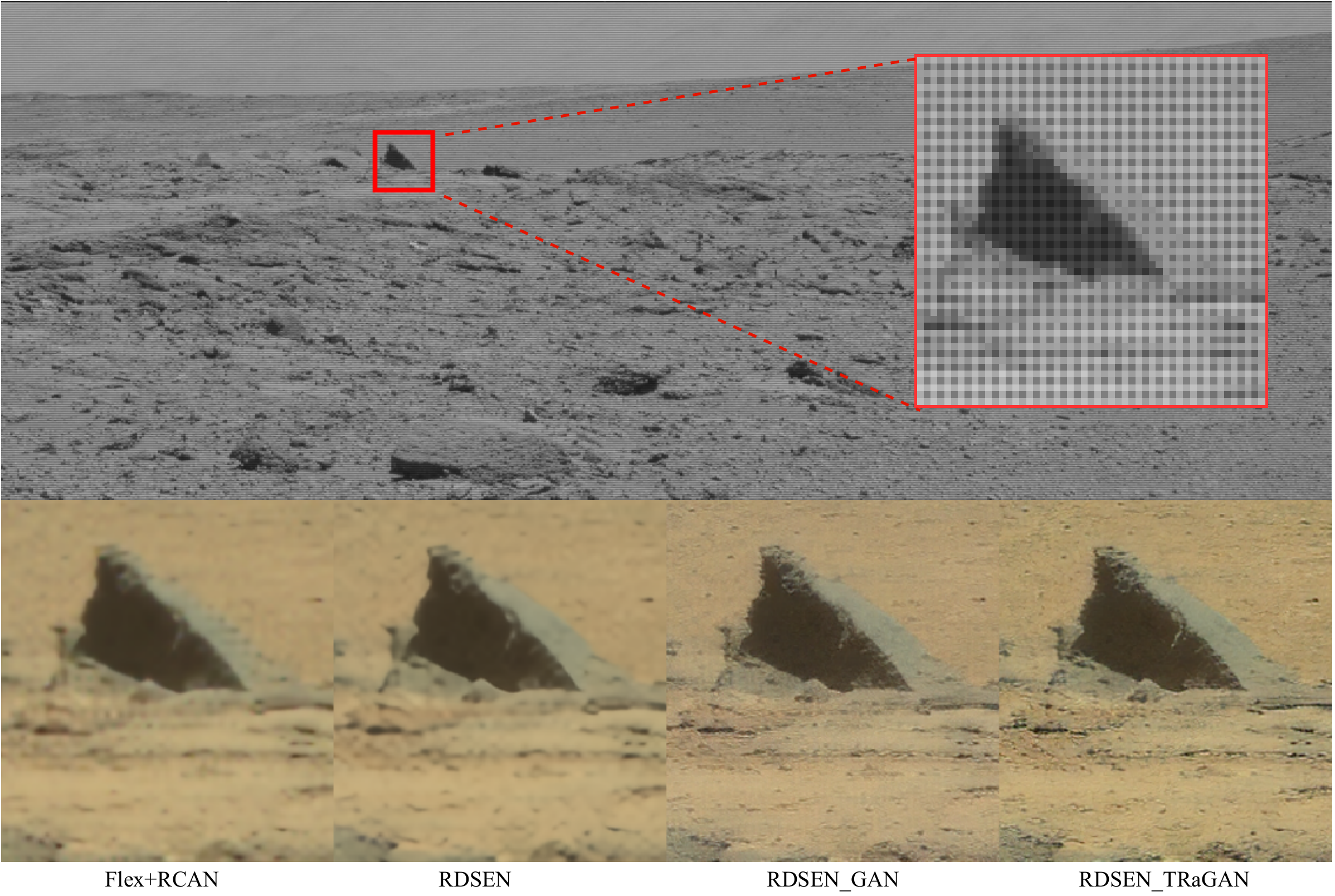}
    \caption{Visual quality comparison of JDSR results on real-world Bayer pattern collected by NASA Mars Curiosity ($4\times$).}
    \label{fig:nasa1}
\end{figure*}

\begin{figure*}
    \centering
    \includegraphics[width=17cm]{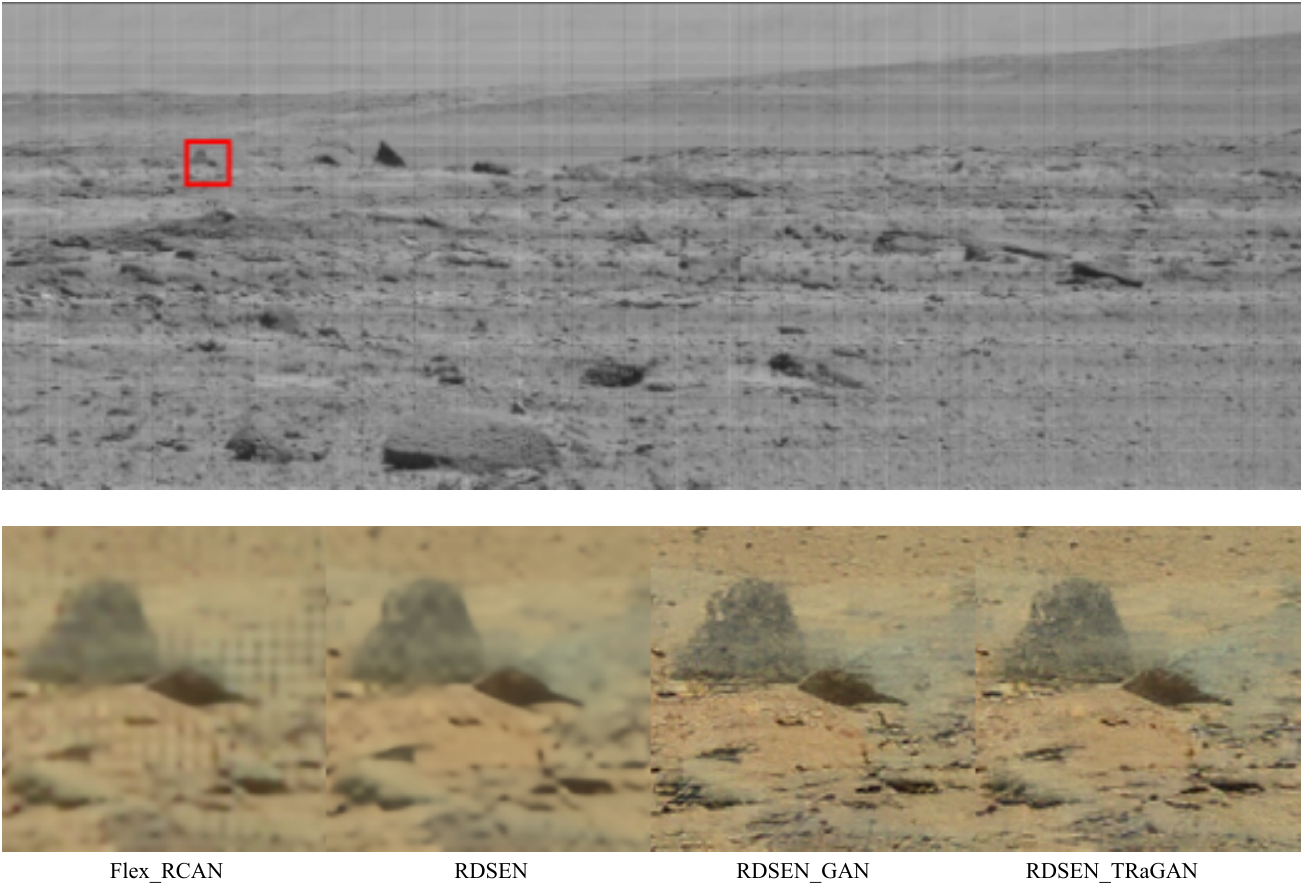}
    \caption{More visual quality comparison of JDSR results on real-world Bayer pattern collected by NASA Mars Curiosity.}
    \label{fig:nasa2}
\end{figure*}

The visual comparison results are shown in Fig.~\ref{fig:nasa1} for a scaling factor of 4. It can be seen that brute-force approach (Flex+RCAN) suffers from undesired artifacts especially around the edge of rocks. Our proposed RDSEN method can overcome this difficulty but the result appears over-smoothed. RDSEN\_GAN improves the visual quality to some degree - e.g., more fine details are present and sharper edges can be observed. Replacing GAN by TRaGAN can further improve the visual quality not only around the textured regions (e.g., roads and rocks) but also in the background (e.g., terrain appears visually clearer and sharper). 
Fig.~\ref{fig:nasa2} shows the visual comparison among Flex+RCAN, RDSEN, RDSEN\_GAN and RDSEN\_TRaGAN approaches. The raw image is captured by the right eye of NASA Mast Camera. The scale factor is 4 (please zoom in to get a better view).

\section{Conclusion}
In this paper, we proposed to study the problem of joint demosaicing and super-resolution (JDSR) - a topic has been underexplored in the literature of deep learning. Our solution consists of a new residual-dense squeeze-and-excitation network for image reconstruction and an improved GAN with relativistic discriminator and new loss functions for texture enhancement. Compared with naive network designs, our proposed network can stack more layers and be trained deeper and wider by newly designed RDSEB block. This is because RDSEB makes multiple residual-dense connection on channel descriptor to allow more faithful information flow. Additionally, we have studied the problem of perceptual optimization for JDSR. Our experimental results have verified that TRaGAN can generate more realistically-looking images (especially around textured regions) and achieve lower PI scores than standard GAN. Finally, we have evaluated our proposed method (RDSEN\_TRaGAN) on real-world Bayer patterns collected by the Mastcam of NASA Mars Curiosity Rover, which supports its superiority to naive network design (e.g., Flex+RCAN) and the effectiveness of perceptual optimization. Another potential application of JDSR in practice is the digital zoom feature in smartphone cameras. Our solution to JDSR offers a cost-effective alternative to the existing hardware-based solutions (e.g. periscope camera design to achieve optical zoom).


%



\section*{Acknowledgment}

The authors would like to thank Dr. Chiman Kwan for supplying real-world Bayer pattern collected by NASA Mars Curiosity. This work is partially supported by the NSF under grants IIS-2027127, IIS-1951504, CNS-1940859, CNS-1946327, CNS-1814825 and OAC-1940855, the DoJ/NIJ under grant NIJ 2018-75-CX-0032, and the WV
Higher Education Policy Commission Grant (HEPC.dsr.18.5).

\ifCLASSOPTIONcaptionsoff
  \newpage
\fi

\end{document}